\documentstyle[12pt]{article}
\def\BorL{b } % portrait mode
\makeatletter
\newif\ifpr@pstyle \pr@pstylefalse
\newif\ifnons@qeq  \nons@qeqfalse
\def\bigpage{
\setlength{\topmargin}{-.5in}
\setlength{\oddsidemargin}{.5pc}
\setlength{\evensidemargin}{.5pc}
\setlength{\textwidth}{35pc}
\setlength{\textheight}{52pc}
\setlength{\parskip}{6pt plus 2pt minus 1pt}
\newlength{\paperbaselineskip}
\setlength{\paperbaselineskip}{20pt plus 0.2pt minus 0.1pt}
\def\@oddfoot{\hfil -- \thepage~--\hfil}
\let\@evenfoot\@oddfoot
        \def\thesection{\arabic{section}.}
        \def\thesubsection{\thesection\arabic{subsection}}
        \def\@ourappendix{\par\setcounter{section}{0}
                      \setcounter{subsection}{0}
                      \def\thesection{\Alph{section}.}
                      \ifnons@qeq
                      \def\theequation{\Alph{section}.\arabic{equation}}\fi}
        \def\appendix{\@ourappendix}
        \def\section{\@startsection {section}{1}%
            {\z@}{5ex plus .2ex minus .4ex}%
            {1.5ex plus.4ex minus .1ex}%
            {\centering\ifpr@pstyle\else\ifx\undefined\reset@font\else%
             \reset@font\fi\large\fi\bf}}
        \def\subsection{\@startsection{subsection}%
            {2}{\z@}{3.25ex plus .4ex minus .4ex}%
            {1ex plus .2ex}{\bf}}
}
\bigpage
\newfont{\fourteencp}{cmcsc10 scaled\magstep2}
\newfont{\titlefont}{cmbx10 scaled\magstep2}
\newfont{\authorfont}{cmcsc10 scaled\magstep1}
\newfont{\fourteenmib}{cmmib10 scaled\magstep2}
\skewchar\fourteenmib='177
\newfont{\elevenmib}{cmmib10 scaled\magstephalf}
\skewchar\elevenmib='177
\newfont{\ninemib}{cmmib9} \skewchar\ninemib='177
\makeatletter
\newcommand\nonsequentialeqnum{
        \nons@qeqtrue
\@addtoreset{equation}{section}
\def\theequation{\arabic{section}.\arabic{equation}}}
\newif\ifp@bblock  \p@bblocktrue
\newcommand\nopubblock{\p@bblockfalse}
\newcommand\topspace{\hrule height 0pt depth 0pt \vskip}
\newcommand\p@bblock{\begingroup \tabskip=\hsize minus \hsize
\baselineskip=1.5\ht\strutbox \topspace-2\baselineskip
\halign to\hsize{\strut ##\hfil\tabskip=0pt\crcr
\the\Pubnum\crcr\the\date\crcr}\endgroup}
\newcommand\YUKAWAmark{\hbox{
        \ifpr@pstyle\ninemib\else\elevenmib\fi
        Yukawa\hskip1mm Institute\hskip1mm Kyoto \hfill}}
\newtoks\date
\newtoks\Pubnum

\newcommand{\frontpageskip}{\vspace{12pt plus .5fil minus 2pt}}
\def\@authoraddress{} \def\@title{}
\def\title#1{\gdef\@title{\frontpageskip
\begin{center}{\titlefont #1}\end{center}\par}}
\def\@author#1{\frontpageskip\par\begin{center}{\authorfont #1}
\end{center}
\nobreak}
\def\author#1{\expandafter\def\expandafter\@authoraddress\expandafter
    {\@authoraddress{\@author{#1}}}}
\def\andauthor#1{\expandafter\def\expandafter\@authoraddress\expandafter
    {\@authoraddress{\frontpageskip\centerline{and}\@author{#1}}}}
\def\authors#1{\expandafter\def\expandafter\@authoraddress\expandafter
    {\@authoraddress{\frontpageskip\noindent #1}}}
\def\@address#1{\par\begin{center}{\sl #1}\end{center}\par}
\def\address#1{\expandafter\def\expandafter\@authoraddress\expandafter
    {\@authoraddress{\@address{#1}}}}
\def\andaddress#1{\expandafter\def\expandafter%
    \@authoraddress\expandafter
    {\@authoraddress{\par\centerline{\sl and}\@address{#1}}}}
\renewcommand{\thanks}[1]{\footnote{#1}}
\def\maketitle{\par
  \begingroup
       \def\thefootnote{\fnsymbol{footnote}}
\thispagestyle{empty}
        \baselineskip=\paperbaselineskip
\@maketitle
\endgroup
\setcounter{footnote}{0}
\let\maketitle\relax \let\@maketitle\relax
\let\@thanks\relax \let\@title\relax
\let\@title\relax \let\@authoraddress\relax
\let\thanks\relax}
\def\@maketitle{%
        \ifpr@pstyle\vspace{-1.0cm}\else\vspace{-1.7cm}\fi
\YUKAWAmark\vskip0.6cm
\ifp@bblock\p@bblock \else\hrule height 0pt \relax \fi
\@title
\@authoraddress
}
\renewcommand{\abstract}{\par\frontpageskip\centerline{
             \ifpr@pstyle\twelvecp\else\fourteencp\fi Abstract}
\vspace{8pt plus 3pt minus 3pt}}
\makeatother
\def\bigmode{b }
\ifx\BorL\undefined\read-1 to\BorL\fi
\makeatletter
\def\doublepage{
        \twocolumn
        
        \pr@pstyletrue
        \sloppy
        \flushbottom
        \setlength{\topmargin}{-0.95in}
        \setlength{\headsep}{20pt}
        \setlength{\headheight}{10pt}
        \hoffset=-0.35in
        \leftmargini 2em
        \leftmarginv .5em
        \leftmarginvi .5em
        \marginparwidth 48pt
        \marginparsep 10pt
        \setlength{\columnsep}{0.7truein}
        \setlength{\textwidth}{10.5truein}
        \setlength{\textheight}{7truein}
        \setlength{\oddsidemargin}{0.0truein}
        \setlength{\evensidemargin}{0.0truein}
        \multiply\paperbaselineskip by 4
                   \divide\paperbaselineskip by 5
        \multiply\footskip by 4 \divide\footskip by 5
        \setlength{\parskip}{4pt plus 1.5pt minus 1pt}
        \newlength{\halfwidth}
        \halfwidth=\textwidth\advance\halfwidth by -\columnsep
                         \divide\halfwidth by 2
        \newfont{\twelvemib}{cmmib10 scaled\magstep1}
                 \skewchar\twelvemib='177
        \newfont{\tenmib}{cmmib10}
                 \skewchar\tenmib='177
        \newfont{\twelvecp}{cmcsc10 scaled\magstep1}
        \def\pagebox{\hbox to \halfwidth{\hfil  -- \thepage~--\hfil}}
        \def\@oddfoot{\pagebox\hfil\addtocounter{page}{1}\pagebox}
        \let\@evenfoot\@oddfoot
        \def\ps@empty{\let\@mkboth\@gobbletwo\let\@oddhead\@empty
               \def\@oddfoot{\hbox to \halfwidth{\hfil ~~~~~~~}\hfil
               \addtocounter{page}{1}\pagebox}
                \let\@evenhead\@empty\let\@evenfoot\@oddfoot}
        \def\appendix{\@ourappendix}
        \def\section{\@startsection {section}{1}%
            {\z@}{5ex plus .2ex minus .4ex}%
            {1.5ex plus.4ex minus .1ex}%
            {\centering\ifpr@pstyle\else\reset@font\large\fi\bf}}
        \def\subsection{\@startsection{subsection}%
            {2}{\z@}{3.25ex plus .4ex minus .4ex}%
            {1ex plus .2ex}{\bf}}
}
\ifx\BorL\bigmode
\else
        \input art10.sty
        \doublepage
\fi
\makeatother
\makeatletter
\newif\ifepsfloaded
\newif\iffigureexists

\ifeof 1 \epsfloadedfalse \else \epsfloadedtrue \fi
\ifepsfloaded \input epsf \fi

\def\checkex#1 {\relax
    \openin 1 #1
    \ifeof 1 \figureexistsfalse
    \else \figureexiststrue
    \fi \closein 1 }
\def\figinsertraw#1#2{
   \ifepsfloaded
       \checkex #1
       \iffigureexists
           \immediate\write16{(#1)}
           #2
       \else
           \immediate\write16{(#1 NOT FOUND!)}
           \vbox to 2in{\hbox to 2in {\hss} \vss}% blank
       \fi
   \else
       \immediate\write16{(NOT inputting #1; no epsf.tex)}
       \vbox to 2in{\hbox to 2in {\hss} \vss}% blank
   \fi}
\newcommand{\reduceland}[2]{\dimen@=#1
     \ifpr@pstyle\multiply\dimen@ by 4\divide\dimen@ by 5\fi
     \edef#2{\dimen@}}
\def\F@gin#1#2#3#4{
  \ifepsfloaded
    \checkex #1
    \iffigureexists
        \immediate\write16{(#1)}
        \begin{figure}
        \ifdim#2>\z@\reduceland{#2}{\dimen@ii}\epsfxsize=\dimen@ii\fi
        \ifdim#3>\z@\reduceland{#3}{\dimen@ii}\epsfysize=\dimen@ii\fi
        \centerline{\epsfbox{#1}}
        {#4} \end{figure}
    \else
        \immediate\write16{(#1 NOT FOUND!)}
        \begin{figure}
        \ifdim#2>\z@\reduceland{#2}{\dimen@ii}\else\dimen@ii=2in\fi
        \ifdim#3>\z@\reduceland{#3}{\dimen255}\else\dimen255=2in\fi
        \centerline{\framebox[\dimen@ii]{\rule{0pt}{\dimen255}#1}}
        {#4} \end{figure}
    \fi
  \else
    \immediate\write16{(NOT inputting #1; no epsf.tex)}
    \begin{figure}
    \centerline{\framebox[2in]{\rule{0pt}{2in}#1}}
    #4\end{figure}
  \fi}
\def\figinsertx#1#2#3{\F@gin{#1}{#2}{0pt}{#3}}
\def\figinserty#1#2#3{\F@gin{#1}{0pt}{#2}{#3}}
\def\figinsert#1#2{\F@gin{#1}{0pt}{0pt}{#2}}
\makeatother
\begin{document}    
\hspace{10cm} hep-th/0111240

\vspace{-0.2cm}
\hspace{10cm} NBI-HE-01-11

\vspace{-0.2cm}
\hspace{10cm} YITP-01-79

\vspace{-0cm}
\hspace{10cm} November 2001

\vspace{2.5cm}
%
%\nopubblock %% uncomment in making submit-version 
%\nonsequentialeqnum %% uncomment in (Sec.Num) eq. number style.
%\pubnum{NBI-HE-01-11, YITP-01-79}
%\date{November 2001}

\title{A New Type of String Field Theory\footnote{To appear in the Proceedings of The 10th International
Symposium on Strings, July 3-7, Fukuoka, Japan, 2001 (AIP Press)}% 
%\thanks{Thank you title}
}

\author{Holger B. Nielsen%
%\thanks{Thank you so much}
}
\address{Niels Bohr Institute, Blegdamsvej 17, Copenhagen \o ~  Denmark}
\andauthor{Masao Ninomiya}
\address{Yukawa Institute for Theoretical Physics, Kyoto University, 
  Kyoto, 606-8502, Japan}
%\andaddress{Another address of the Collaborator}

\maketitle

\begin{abstract}
We propose a new way of second quantizing string theory.
The method is based on considering the Fock space of strings described 
by constituents which make up the $X^\mu_R$ and the $X^\mu_L$ i.e. the right
and left mover modes separately.
A state with any number of strings get represented by the Cartesian
product of two free particle Fock spaces, one for right mover degrees
of freedom, and one for left.
The resulting string field theory is a free theory.  

\end{abstract}

\newpage
\section{Introduction}
There exist already several variants of string field theories along the 
line of the Kaku-Kikkawa's one\cite{kaku} which has for any state 
single-string creation and annihilation operators so that various numbers of
strings can be present in the different single-string states.
In models of this kind of second quantized string theories one can
distinguish two two-string states which are denoted as $|1\rangle$ and
$|2)\rangle$ (see Figure 1) although they look somewhat
similar in the following way:

1) The state $|1)\rangle$ is a two-particle state in which two open strings
are present 
in such a configuration that the two strings lie just along the same
curve for a piece somewhere in the middle of the strings.

2) The state $|2)\rangle$ is a corresponding two-string state to the one
under 1), but the two strings follow each other somewhere in the
middle by permuting so to speak the ``tails'' of the 
two strings in the Fock-space state $|1)\rangle$.
That is to say that the two-particle state $|2)\rangle$ in the Fock space
describes two strings one of which ``half'' coincides
with a piece of string number one in Fock state  $|1)\rangle$ while the 
other ``half'' instead coincides with the ``tail'' part of the
second string in Fock state $|1)\rangle$.

The two Fock states $|1)\rangle$ and $|2)\rangle$ have some string material
present -- in single or double amounts -- in just the same curve
pieces in space, so that they can only be distinguished if one can
find out how the string pieces hang together.
Nevertheless string field theories such as Kaku-Kikkawa's
one\cite{kaku}, Kyoto group's (HIKKO's) one\cite{hata}, 
Witten's cubic one\cite{witten}, and Zwiebach's one\cite{zwiebach}
have two-particle states $|1)\rangle$
and $|2)\rangle$ as mentioned that are counted as quite different,
distinguishable Fock space states.

\begin{center}
\figinsertx{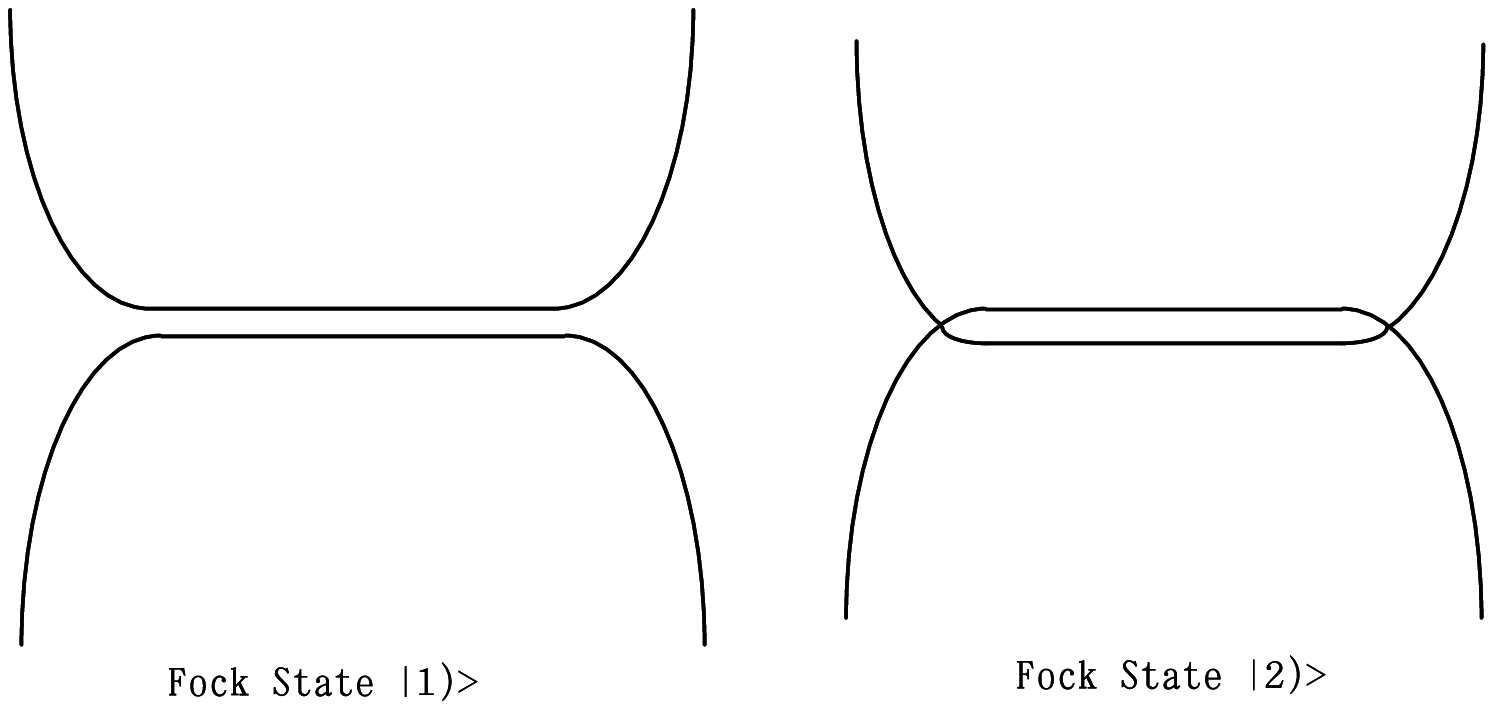}{20pc}{}
\end{center}

\vspace{-1.7cm}
It is the purpose of the present work to present ideas to make a
string field theory model representing the class of 
theories in which the Fock states $|1)\rangle$ and $|2)\rangle$ are
not distinguishable but rather represent the same physical
state.
This class of models has not been much studied unless one counts that
the strings of QCD as well as the strings of matrix
models\cite{banks, ishibashi} are
really of this type.

In QCD you have only local fields to describe where a string
is present and it would seem very hard to see how two
QCD-strings lying on top of each other along a piece of curve could
get their ``heads'' and ``tails'' associated with each other by
QCD degrees of freedom information.
So it seems much more likely that QCD develops the type of
strings where the states $|1)\rangle$ and $|2)\rangle$ just
described above must be identified.
This conclusion becomes even more obvious if we use a strong coupling
approximation as the method for implementing the strings into the
lattice QCD or just Yang Mills theories.  
Then, the strings become flux quanta of color electric flux
and there is no way to keep track of or identify parts of the same
string. 

QCD or Yang Mills theories as well as also matrix models provides models of 
string field theories of the same type as we are going to propose in the
present article.

It is, however, our goal to make string field theory model that does
not need a very hard and non-linear calculation to connect to the 
string picture as QCD needs.

We shall indeed see that our model is inspired by an infinite set of
seemingly conserved quantities noticeable in classical (i.e. non
quantum mechanical) string theory, as we shall explain in the
following section II.
Then we shall start the description of our string field theory in
section III.
A crucial complication of our model is that it needs a constraint
ensuring that each ``constituent'' in $X_R$- or $X_L$-space has a successor
constituent as shall be described in section IV.
Since our model has at first some bad features because too many 
states have been made identical it is far from obvious that our model
is indeed an acceptable string field theory.
It is therefore absolutely crucial that it
could be used to deduce the Veneziano-model scattering amplitude.
That we shall briefly sketch in section V.
Finally in section VI we shall resume and conclude among other things
that our model is a free theory and that it thus becomes important for 
judging the validity of string theory as a model for nature, if really 
a free theory could be the model for nature.

\section{Inspiration by the conservation of right and left moving patterns}

The crucial observation that has inspired our proposal for string
field theory originates from considering classical string
``scattering'' which takes place by a couple of permuting their
``tails'' when strings touch in one point.  
Hereby we understand that, say two strings come along in such a way
that in a moment of time they have one point in common, but that then
after this moment the strings develop as if they were a different pair 
of strings.  Namely the one obtained by combining the first part of the 
string number 1 with the second part of string number 2, and vice
versa.  
We call that the strings get their tails permuted when the beginnings
and ends of the original strings are
combined with the ends in an different way.

Consider -- in classical approximation -- two strings described before 
the collision by:

The first string:

\begin{eqnarray}
X^\mu_I (\sigma, \tau)=X^\mu_{RI}(\sigma, \tau)+X^\mu_{LI}(\sigma, \tau)\nonumber\\
=X^\mu_{RI}(\tau-\sigma)+X^\mu_{LI}(\tau+\sigma)
\end{eqnarray}

The second string:

\begin{equation}
X^\mu_{II}(\hat{\sigma}, \hat{\tau})=X^\mu_{RII}
(\hat{\tau}-\hat{\sigma})-X^\mu_{LII}(\hat{\tau}+\hat{\sigma})
\end{equation}

It is important that the two strings are described with conformal
gauge choice and with Minkowskian metric.  
This gauge choice does not fix the gauge (of reparametrization) freedom
completely, but only poses the restrictions.  

\begin{equation}
\frac{\partial X^\mu_I}{\partial\tau}\frac{\partial X_{I\mu}}{\partial\tau}=-\frac{\partial
X^\mu_I}{\partial\sigma}\frac{\partial X_{I\mu}}{\partial\sigma}
\end{equation}

and 

\begin{equation}
\frac{\partial X^\mu_I}{\partial\sigma}\frac{\partial X_{I\mu}}{\partial \tau}=0
\end{equation}

and analogous ones for string II, i.e. for $X^\mu_{II}$ instead of
$X^\mu_I$.  
With this reparametrization gauge choice the equations of motions
become

\begin{eqnarray}
\left(\frac{\partial^2}{\partial\tau^2}-\frac{\partial^2}{\partial\sigma^2}\right)X^\mu_I=0\\
\left(\frac{\partial^2}{\partial\hat{\tau^2}}-\frac{\partial^2}{\partial\hat{\sigma^2}}\right)X^\mu_{II}=0
\end{eqnarray}

and it is these equations that are solved by writing

\begin{eqnarray}
X^\mu_I(\sigma, \tau)&=&X^\mu_{RI}(\tau-\sigma)+X^\mu_{LI}(\tau+\sigma)\\
X^\mu_{II}(\hat{\sigma},
\hat{\tau})&=&X^\mu_{RII}(\hat{\tau}-\hat{\sigma})+X^\mu_{LII}(\hat{\tau}+\hat{\sigma})
\end{eqnarray}

In fact will any 26 pairs of functions $X^\mu_{RI}$ and $X^\mu_{LI}$
only depending on $\tau_R=\tau-\sigma$ and
$\tau_L=\tau+\sigma$ respectively lead to solution of the equation of motion.

When at a moment string I and string II have a common point it means
that there exist two sets of timetrack surface coordinates ($\sigma_o,
\tau_o$) and ($\hat{\sigma_o}, \hat{\tau_o}$) such that 

\begin{equation}  
X^\mu_I(\sigma_o, \tau_o)=X^\mu_{II}(\hat{\sigma_o}, \hat{\tau_o})
\end{equation}  

After the collision we imagine that there are in fact two new strings
which we may call III and IV composed from pieces of the original
strings I and II and that the development goes on as if III and IV are 
the strings then.  
Formally the relation -- and we here think locally at first ignore for 
simplicity the problems of boundary conditions -- among the strings 
III and IV to I and II are simply at $\tau=\tau_o$ and
$\hat{\tau}=\hat{\tau_o}$ described as
$\tau_{III}=\tau_{IIIo}=\tau_o$ moment for string III and
$\tau_{IV}=\tau_{IVo}=\tau_o$ too say.

\begin{eqnarray} 
X^\mu_{III}(\sigma_{III}, \tau_{IIIo})=\left\{
\begin{array}{l}
X^\mu_I(\sigma_{III},
\tau_o) ~~ {\mathrm for} ~~ \sigma_{III}>\sigma_o\\
X^\mu_{II}(\sigma_{III}-\sigma_o+\hat{\sigma_o}, \hat{\tau_o}) ~~
{\rm for} ~~ \sigma_{III}<\sigma_o
\end{array}
\right.
\end{eqnarray} 

\begin{eqnarray} 
X^\mu_{IV}(\sigma_{IV}, \tau_{IVo})=\left\{
\begin{array}{l}
X^\mu_{II}(\sigma_{IV}-\sigma_o+\hat{\sigma_o}, \hat{\tau_o}) ~~
{\rm for} ~~ \sigma_{IV}>\sigma_o\\
X^\mu_{I}(\sigma_{IV},
\tau_o) ~~ {\rm for} ~~ \sigma_{IV}<\sigma_o
\end{array}
\right.
\label{scatterrel}
\end{eqnarray} 

Also the $\tau_{III}$ and $\tau_{IV}$ derivatives obey the
analogous relations; you just take the $\tau$-derivatives.
Using both equations (\ref{scatterrel}) and the corresponding
$\frac{\partial}{\partial\tau}$ relations we have information enough 
to put the solutions for the development of the strings III and IV in
terms of the left and right mover functions from the strings I and
II. 
Indeed we may even simply argue that from causality at finite distance
from the point of collision $X^\mu_I(\sigma_o,
\tau_o)=X^\mu_{II}(\hat{\sigma_o},
\hat{\tau_o})=X^\mu_{III}(\sigma_o, \tau_o)=X^\mu_{IV}(\sigma_o,
\tau_o)$ tight in $\tau$ to the ``moment of collision'' in
$\tau_{III}, \tau_{IV}$ etc. compared to the $\sigma$-distance to
the collision point solutions in string III and string IV must be
identical to the corresponding ones in I and II locally.
What goes on has simply no knowledge of whether the collision took
place.  
Hence the solutions for all later ``time'' or better ``$\tau$'' are:

For string III with $\tau_{III}\geq\tau_{IIIo}=\tau_o$

\begin{equation}
X^\mu_{III}(\sigma_{III}, \tau_{III})=X^\mu_{RIII}(\tau_{III}-\sigma_{III})+X^\mu_{LIII}(\tau_{III}+\sigma_{III})
\end{equation}

where

\begin{eqnarray}
X^\mu_{RIII}(\tau_{III}-\sigma_{III})=\left\{
\begin{array}{l}
X^\mu_{RI}(\tau_{III}-\sigma_{III}) ~~
{\rm for} ~~ \tau_{III}-\sigma_{III}>\tau_o-\sigma_o\\
X^\mu_{RII}(\tau_{III}-\sigma_{III}-\tau_o+\sigma_o+\hat{\tau_o}-\hat{\sigma_o})
~~ {\rm for} ~~ \tau_{III}-\sigma_{III}<\tau_o-\sigma_o
\end{array}
\right.
\end{eqnarray}

and 

\begin{eqnarray}
X^\mu_{LIII}(\tau_{III}-\sigma_{III})=\left\{
\begin{array}{l}
X^\mu_{LI}(\tau_{III}+\sigma_{III}) ~~
{\rm for} ~~ \tau_{III}+\sigma_{III}<\tau_o+\sigma_o\\
X^\mu_{LII}(\tau_{III}+\sigma_{III}-\tau_o+\hat{\tau_o}-\sigma_o+\hat{\sigma_o})
~~ {\rm for} ~~ \tau_{III}+\sigma_{III}>\tau_o+\sigma_o
\end{array}
\right.
\end{eqnarray}

For string IV:

\begin{equation}
X^\mu_{IV}(\sigma_{IV}, \tau_{IV})=X^\mu_{RIV}(\tau_{IV}-\sigma_{IV})+X^\mu_{LIV}(\tau_{IV}+\sigma_{IV})
\end{equation}

where

\begin{eqnarray}
X^\mu_{RIV}(\tau_{IV}-\sigma_{IV})=\left\{
\begin{array}{l}
X^\mu_{RII}(\tau_{IV}-\sigma_{IV}-\tau_o+\sigma_o+\hat{\tau_o}-\hat{\sigma_o}) ~~
{\rm for} ~~ \tau_{IV}-\sigma_{IV}>\tau_o-\sigma_o\\
X^\mu_{RI}(\tau_{IV}-\sigma_{IV})
~~ {\rm for} ~~ \tau_{IV}-\sigma_{IV}<\tau_o-\sigma_o
\end{array}
\right.
\end{eqnarray}

and 

\begin{eqnarray}
X^\mu_{LIV}(\tau_{IV}+\sigma_{IV})=
\left\{ 
\begin{array}{l}
X^\mu_{LII}(\tau_{IV}+\sigma_{IV}-\tau_o-\sigma_o+\hat{\tau_o}+\hat{\sigma_o}) ~~
{\rm for} ~~ \tau_{IV}+\sigma_{IV}<\tau_o+\sigma_o\\
X^\mu_{LI}(\tau_{IV}+\sigma_{IV})
~~ {\rm for} ~~ \tau_{IV}+\sigma_{IV}>\tau_o+\sigma_o
\end{array}
\right.
\end{eqnarray}

Now the observation which is so important for inspiring the proposed
attempt in this article to second quantize string theory is the
following:

On string III and IV together you will find realization of any value
of $X^\mu_{RIII}$ and $X^\mu_{RIV}$ and any pattern just once for each 
time; you find such value or pattern for the right mover position on I
and II together i.e. on $X^\mu_{IR}$ and $X^\mu_{IIR}$.  
The same result follows quite analogously for the $X_L$'s.

It should be noted that since only $X_R+X_L$ or $\dot X_R, \dot X_L$ 
have physical meaning you always make a kind of gauge transformation
by making the transformation

\begin{eqnarray}
\label{transf}
X^\mu_R\rightarrow X^\mu_R+k\\\nonumber
X^\mu_L\rightarrow X^\mu_L-k
\end{eqnarray}

As we formulated our observation just now it is only true by an
appropriate adjustment of this freedom in values.
The solutions which we just proposed were, however, put with the
choice that made our observation work as stated.
If we prefer to state our observation in a way not suffering from
this need for adjustment of notation of $X^\mu_R$ and $X^\mu_L$ we may
state it for the derivatives with respect to $\tau_R=\tau-\sigma$ and
$\tau_L=\tau+\sigma$ respectively instead:

The strings III and IV contain together a value-spectrum for their
$\frac{dX^\mu_{RIII}}{d\tau_{RIII}}$ and
$\frac{dX^\mu_{RIV}}{d\tau_{RIV}}$ which is just the same as that
for the incoming strings I and II, i.e. for their 
$\frac{dX^\mu_{RI}}{d\tau_{RI}}$ and
$\frac{dX^\mu_{RII}}{d\tau_{RII}}$. 

More formally stated we may formulate this observation in the following:

%\begin{eqnarray}
%\left\{
%\begin{array}{c|c}
%\dot X^\mu_{RIII}(\tau_{RIII})&\tau_{RIII} ~ \mathrm{a} ~
%\tau_{RIII}=\tau_{III}-\sigma_{III}\\ 
%&{\rm ~ value ~ realized ~at ~ ``some} ~\\
%&\mathrm{later ~ moment'' ~ on ~ string ~ III}
%\end{array}
%\right\}\\\nonumber
%U\left\{\dot X_{RIV}(\tau_{RIV})|\tau_{RIV} ~ \mathrm{a} ~
%\tau_{RIV}=\tau_{IV}-\sigma_{IV} {\rm ~ value ~ realized ~
%at ~ ``some ~
%later ~ moment'' ~ on ~ string ~ IV}
%\right\}\\\nonumber
%=\left\{\dot X_{I}(\tau_{R})|\tau_{R}=\tau-\sigma 
%{\rm ~ a ~ value ~ realized ~
%at ~ ``some ~
%later ~ moment'' ~ on ~ string ~ I}
%\right\}\\\nonumber
%U\left\{\dot
%X_{RII}(\hat{\tau}_{R})|\hat\tau_{R}=\hat\tau-\hat\sigma 
%{\rm ~ a ~ value ~ realized ~
%at ~ ``some ~
%later ~ moment'' ~ on ~ string ~ II}
%\right\}
%\end{eqnarray}

\begin{eqnarray}
\left\{
\dot X^\mu_{RIII}(\tau_{RIII})|\tau_{RIII} ~ a ~ %
\tau_{RIII}=\tau_{III}-\sigma_{III} \mbox{\rm ~ value realized
at ``some later moment'' on string III}
\right\}\\\nonumber
U\left\{\dot X_{RIV}(\tau_{RIV})|\tau_{RIV} ~ a ~ %
\tau_{RIV}=\tau_{IV}-\sigma_{IV} \mbox{\rm ~ value realized
at ``some
later moment'' on string IV}
\right\}\\\nonumber
=\left\{\dot X_{RI}(\tau_{R})|\tau_{R}=\tau-\sigma 
\mbox{\rm ~ a value realized
at ``some 
earlier moment'' on string I}
\right\}\\\nonumber
U\left\{\dot
X_{RII}(\hat{\tau}_{R})|\hat\tau_{R}=\hat\tau-\hat\sigma 
\mbox{~ a value realized 
at ``some
earlier moment'' on string II}
\right\}
\end{eqnarray}

Of course we have the analogous result for the $X^\mu_L$'s.

Also this result generalizes to the case of successive scatterings of
the type described  -- i.e. a common point at one moment and a
tail exchange -- and we may loosely state the general result:

Any piece of pattern of the $\dot X^\mu_R$-values ( or with
appropriate adjustments of $X^\mu_R$ itself) in the ``incoming'' set of
strings will reappear just once on the $X^\mu_R$'s of the outgoing
strings.
In other words such $\dot X^\mu_R$ (or $X^\mu_R$) patterns are
conserved objects.

In this formulation we had in mind an S-matrix-like situation of
classical strings, i.e. a set of classically treated strings come in
from far way and scatter by what locally looks like hit in one point
with tail-exchange.  
Also the strings are imagined to separate infinitely at the end so
that we are allowed to use the concepts of ``incoming'' and
``outgoing'' strings.

The formal formulation of this general version of the observation is 

\begin{eqnarray}
U_{i\epsilon\left\{\mbox{\footnotesize ``incoming'' strings}\right\}}
\left\{
\dot X^\mu_{Ri}(\tau_{Ri})|\tau_{Ri}=\tau_i-\sigma_i\mbox{ ~ an early 
realized $\tau_{Ri}$-value on string i}
\right\}\\\nonumber
=U_{j\epsilon\left\{\mbox{\footnotesize ``outgoing'' strings}\right\}}
\left\{
\dot X^\mu_{Rj}(\tau_{Rj})|\tau_{Rj}=\tau_j-\sigma_j\mbox{ ~ a late 
realized $\tau_{Rj}$-value on string j}
\right\}
\end{eqnarray}

It should be admitted that strictly speaking there is a lack of
proving the observation for these very values in $\tau_{Ri}$ and
$\tau_{Rj}$ which correspond to the hit-points -- the common points
for crossing strings.  
So strictly speaking the statement is only valid modulo this
supposedly measure null set of hit-points.
By continuity of the functions it should not matter so much though. 

Really the theorem as stated is only true for a theory with only
closed strings in which case we have also the analogous one 
for left mover i.e. $X^\mu_L$'s.  
But  $X^\mu_R$-waves can run to the end of the string and we now want
to ensure and remind the reader that the pattern in $X^\mu_R$ is at
the end of the string reflected as an
$X^\mu_L$-pattern of same sort.

Indeed let us remember the usual boundary condition at the end of the
string -- let us say that $\sigma$ there is 0,

\begin{equation}
\frac{\partial}{\partial\sigma}X^{\mu}(0, \tau)=X^{\mu'}(\sigma=0, \tau)=0
\end{equation}

which implies

\begin{equation}
\dot X^\mu_R(\tau_R=\tau)-\dot X^\mu_L(\tau_L=\tau)=0
\end{equation}

This equation must hold for all $\tau$ and thus $X^\mu_R(\tau)$
can only deviate by an additive constant

\begin{equation}
X^\mu_R(\tau)=X^\mu_L(\tau)+\mathrm{const}
\end{equation}

from $X^\mu_L(\tau)$.  
So really since this constant could be shuffled away by a
transformation of the type (\ref{transf}) we can simply say that :

For open string models we can take 

\begin{equation}
X^\mu_L=X^\mu_R
\label{RLens}
\end{equation}

The boundary constraint at the other end, where according to the
usual convention $\sigma|_{\mbox{\footnotesize second boundary}}=\pi$ we get rather 

\begin{equation}
X^\mu_R(\tau-\pi)=X^\mu_L(\tau+\pi) \mathrm{~~~~~~~~~~~~~ (modulo ~  constant)}
\end{equation}

which together with (\ref{RLens}) leads to the requirement of
periodicity of $X^\mu_R=X^\mu_L$ as a function of the argument,
$\tau$ (up to an additive constant).  

Actually even in closed string case where $X^\mu_R$ and $X^\mu_L$ are
not connected they have to be periodic (up to an additive
constant), for the string to close as a circle.   
This is because there shall be periodicity with respect to $\sigma$ (for fixed
$\tau$) and say $\tau_R=\tau-\sigma$ so that periodicity
with respect to $\tau_R$ is also needed.
 
Our formulated observations for closed string above will for theories
involving also open strings instead be:

Patters -- or say simply $\dot X^\mu_R$ and $\dot X^\mu_L$ values --
found on the ``incoming'' strings $X^\mu_L$ or $X^\mu_R$ (all counted 
together) will reappear just once each on the combined set of
``outgoing'' strings counting for them both $X^\mu_R$ and 
$X^\mu_L$.  

The inspiration to make our string field theory from this conservation
of patterns on $X^\mu_R$ and $X^\mu_L$ is the following: If we represent --
as we do in our model -- the $X^\mu_R$ or
$X^\mu_L$ values taken on by what we call ``constituents''
placed at those points in the $X^\mu_L$- or $X^\mu_R$-spaces, then
these ``constituents'' sit at quite the same places in correspondence
to the incoming set of strings as corresponding to the outgoing set.
In other words, although scatterings as described goes on constituents
representing $X^\mu_R$ or $X^\mu_L$ do not change their ``position'' say.   
In the case of open string
theories the $X^\mu_R$- and $X^\mu_L$-spaces are combined to one common
space.
These constituents does not do anything.
They are just sitting undisturbed and changing neither position nor
momentum.

So at least we have a timeless description telling a lot of
information about the states and developments in a string theory with
classically treated strings if we know the ``constituents'' in
$X^\mu_R$- and $X^\mu_L$-spaces in the closed case, or the
combined $X^\mu_L$- and $X^\mu_R$-space in the also open string case.

Although it is actually the point of view of our string field theory
to throw away as only imagination all other information than
that of the just introduced constituents, it must be admitted that
there is at least some information about the strings which is not
described even if one get to know the positions (and momenta) of all
the constituents in the $X^\mu_R$ etc. 
This lacking information includes at least the information about how
the different pieces of strings hang together. 

For example we stressed above that the $X^\mu_R$ and $X^\mu_L$
patterns found, and thus the constituents representing them would be
quite the same if we just had string I and string II as if we had
instead string III and string IV say moving undisturbed at all times.  
These two thinkable string developments a)I+II happening not to
interact and b)III+IV also happening not to interact, would have quite 
the same combined patterns or constituents. 
They would therefore be quite undistinguishable if one has no further
markings to distinguish these two situations, as shall in fact be seen 
to be the case in our string field theory.  
The reader may have good reason to worry if we are throwing away too
much information, since after all one would expect that it ought to
make sense to distinguish I+II from III+IV existing without scattering.
Actually it is even worse, since only knowing the constituents would
also not distinguish the two just mentioned developments, I+II and
III+IV from the scattering development I+II$\rightarrow$III+IV nor
from III+IV$\rightarrow$I+II.

\section{Setting up our string field theory}

One of the basic ideas of our string field theory is to construct an
actually timeless (i.e. Heisenberg picture) Fock space or Fock space
analogue geared to describe -- only! -- the conserved patterns or
image curves in the $X^\mu_R$- and $X^\mu_L$-spaces for classically treated
strings.  
Obviously the sets of values taken on in these $X^\mu_R$- and
$X^\mu_L$-spaces are (continuous) curves, rather than discrete points, 
since the $X^\mu_R(\tau_R)$ and $X^\mu_L(\tau_L)$ are functions of one 
variable $\tau_R=\tau-\sigma$ or $\tau_L=\tau+\sigma$ respectively.
Nevertheless we are allowed as our special formulation or
model to let them be represented by a very dense chain of point
positioned constituents.  
In the next section we shall go a bit more into a rather detailed
constraint which we shall propose that these
constituents must indeed lie in long chains, thereby to some extend
enforcing the one dimensional curve nature of the constituent chains.
We shall use the same constraint to impose that crudely speaking
$X^\mu_R$ only carry half a degree of freedom in as far as $X^\mu_L$
carry the other half of the original $X^\mu=X^\mu_L+X^\mu_R$.
A priori we shall just make creation and annihilation operators for a
Fock space filling in or removing ``constituents'' in the say
$X^\mu_R$-space. 
Concerning the details of the construction of this Fock space it will
turn out to be a more familiar task if we first notice that the constraints
on the $X^\mu(\sigma,\tau)$ in the single string formulation $\dot
X^\mu\dot X_\mu-X'^{\mu}X'_\mu\approx0$ and $\dot X^\mu X'_\mu\approx0$
to be ``weakly implemented'' in terms of the $X^\mu_R$ and $X^\mu_L$
become $\dot X^\mu_R\dot X_{R\mu}\approx0$ and $\dot X^\mu_L\dot
X_{L\mu}\approx0$.  Or if we think of $\dot X^\mu$ as a 26-momentum density 
then the 26-momentum densities function as the 26-momentum for massless 
onshell particles.
Part of the momentum density of the single string will be in $\dot
X^\mu_R$ part in $\dot X^\mu_L$; When we go to the constituents it
would be the natural suggestion that integrating say $\dot X^\mu_R$
over the bit of $\tau_R$ corresponding to that ``constituent'' should
give the 26-momentum of this constituent.
With such an interpretation into the ``constituent'' language the
condition of constraint $(\dot X^\mu_R)^2\approx0$ (as ``weak
constraint'') becomes the onshell condition (i.e. equation of motion)
for the constituent.
So if we make the Fock space for the constituents in
$X^\mu_R$-space a completely usual one with only onshell particles
possible to create or to annihilate should be applicable.
The restriction to onshell should only imply that the constraint $(\dot 
X^\mu_R)^2\approx0$ gets ensured at the end.

We are now prepared for the set up of our string field theory model in 
a couple of steps: 

1) In the first step we set up the simple and usual Fock space with
constituents which are able at first also not to sit in chains
(essentially curves with the only half a degree of freedom).

2) In the next step, we impose the constraints so that only
``chains'' of constituents can actually be created at a time.

At this step ends in principle the set up of the Fock space, but it
must be admitted that although the translation of the one or two Fock
spaces into strings is relatively simple, it is not at all
obvious that one would get the idea of interpreting the model that way 
if we really happened to live in such a world.
So we rather strongly need a third step in explaining our model as a
string theory:

3) Interpretation of the one or two Fock space model as a string
(field) theory, by interpreting sums of $X^\mu_R$'s (and $X^\mu_L$'s) for 
two constituents as meaning that a string passes the space time at the 
event with this sum as coordinates.

\subsection{Steps of setting up the model}

\subsubsection{Particle Fock space}

1) The basis for our string field theory is in the case of the only
closed string model the Cartesian product of two Fock spaces, $H_R$ and 
$H_L$ each of which is simply the ordinary particle Fock space for
massless free -- scalar in the case of the bosonic 26-dimensional string
theory -- particles that can be created and annihilated into
all the onshell states of a single massless scalar.

One point to be thought about is that this scalar particle shall in
fact like a $\pi^\circ$-meson be its own antiparticle (i.e. the
analogue of a majorana particle for fermion case).  
One indication for this is that the strings have no charge
proportional to the length.

It means that the creation and annihilation operators are defined only 
over positive $\dot X^o_R$ states while the ones with $\dot X^o_R$
negative are related to the ones with positive -- by hermitian
conjugation --. 

2) Second step is that we impose a constraint telling that if we have
one constituent we have also a ``successor'' in a single
particle state that is obtained from the first one by action with a certain
operator 

\begin{equation}
\exp\left(i2\pi\alpha'p^2+if(X)\right)
\label{suc}
\end{equation}

This is to be understood that if there is a particle $A$ in the single
particle state $\psi_A$ there must be its successor $B$ in the state
$\psi_B$ obtained from $\psi_A$ by 

\begin{equation}
\psi_B=\exp\left(i2\pi\alpha'p^2+if(X)\right)\psi_A
\end{equation}

In this operator $X^\mu$ and $p^\mu$ are the position and momentum
operators for the $\mu$-th coordinate and $p^2=p^\mu p_\mu$.

The function $f$ can be chosen.  
Therefore there is a freedom to choose the operator by choosing
$f$ differently, all the time obtaining a satisfactory successor $B$.
If there were no such freedom in the successor producing operator
(\ref{suc}) the successor $B$ would be totally determined by $A$ and
there would be no way of having different chains with the same starting 
constituent. 
It should, however, be noted that $f$ is just one real function and
that there is no corresponding function to vary depending on $p^\mu$,
while a constituent $B$ in a wave packet would in analogy to classical 
physics expectedly be possible to change a bit compared to the
foregoing constituent $A$ by two free parameters per dimension.
We imagine to take $f(X)$ linear $f(X)=f_\mu X^\mu$ so that it is only 
one parameter per dimension.
It is this limited amount of freedom in setting up the successors that 
we refer to by saying that the chain of constituents only corresponds 
``half a degree of freedom'' -- You can adjust the momentum of the
successor $B$ by adjusting $f$ or $f^\mu$, but you cannot adjust its
position (directly).

The reason for the specific form, and especially the funny appearance
of the parameter in string theory $\alpha'$ will be postponed to next
section.  

To make more precise the meaning of the requirement of having
successors to all constituents it may be best to describe the allowed
subspace $H_{R \; \; \mbox{\footnotesize allowed}}$.  It consists of 
those Fock states,
i.e. states from $H_R$ for which all the particles have their
appropriate successor that is produced from a vacuum state by acting with
products of creation operators, creating whole chains of constituents.

We may in fact think of a product of creation operators in 
which each factor is some creation operator creating a particle/constituent in
some wave packet like state $\psi_i$. 
Then the product 

\begin{equation}
\displaystyle\Pi_{i\epsilon\mbox{\footnotesize``chain''}} a^+(\psi_i)
\end{equation}

is defined to be an allowed chain of creating operators, provided that the
series of single particle states $\psi_o, \psi_1, \psi_2, \cdot\cdot\cdot,
\psi_i, \cdot\cdot\cdot$ (presumably a closed chain, or a to both sides infinite
one) obey

\begin{equation}
\psi_{i+1}=\exp\left(i2\pi\alpha'p^2+if^{(i)}_\mu X^\mu\right)\psi_i
\end{equation}

for all $i$ counted cyclically for the case of a closed chain.

3) Third step is the interpretation of the model into a string
language.

As the model in step 1) and 2) is set up we have constructed
constituent chains in the $X^\mu_R$- and $X^\mu_L$-space (or in the
combined one in open string model case) which are of dimension 1 in spaces 
with same dimension as space time, in the well working bosonic string
case $d=26$. 
Thus the chains defined under 2) are in the high
density limit of constituents similar in dimension as time tracks of
particles a priori.  (it must though be contemplated that the onshell
condition from 1) actually enforces them to be in a superposition
extended infinitely in some direction.)  

The strings, i.e. the string time tracks which are 2-dimensional
embeddings into 26-dimensions, come out by asking for the set 

\begin{equation}
\left\{
X^\mu_R(A)+X^\mu_L(B)|A, B ~ \mathrm{constituents}
\right\}
\end{equation}

which become two-dimensional, once the chains in $X^\mu_R$- and
$X^\mu_L$-spaces are one dimensional tracks.

This interpretation actually represents a worrisome point for the
model of ours because we have no information in our formalism telling
which chains to combine with which.
So a priori we obtain string-time-tracks from all possible combinations 
of one curve in $X^\mu_R$-space with one in $X^\mu_L$-space.
This is what we call the ``cross combination problem'' of our model
and it is clearly not a property of a good physical model and it is not
contained in conventional string theory either.  

The hope that this ``problem'' is not really a problem may run like
this:

Physicists living in a world of strings would typically make (thought) 
experiments of the sort that they arrange or find out the state of
some strings and then some time may pass and they look for another set 
of strings and ask the theory for the probability (density) for
such happening.
This type of experiment is really the $S$-matrix or an approximate
$S$-matrix type experiment.

In our string field theory the knowledge about the ``incoming''
strings -- the ones in the initial state -- will mean that we have to
have those chains or curves of constituents in the $X^\mu_R$- and
$X^\mu_L$-spaces that can give these strings.

As we argued in the classical approximation in the second section the
$X^\mu_R$ and $X^\mu_L$ patterns are conserved so that the outgoing
and the ingoing string systems are indeed sets of strings not
distinguishable if one only keeps the information of the constituent
chains as in our string field theory.
So typically the states with the constituents and their chains which split
up and combine in a new way corresponds to what could be scattering
results of the strings.
So one might hope that finding the cross combined strings could be
interpreted as seeing the scattering of strings.

It must be admitted though that this hope may not quite work in the
case of two closed strings which have simply their $X^\mu_L$-mover
degrees of freedom permuted. 
In fact but there is a somewhat lucky occurrence of the common points of the
strings: It could be that the strings in the final state classically just 
could turn out to be the same as the one obtained by such permutation
of the $X^\mu_L$-degrees of freedom.
So should one really observe such a pair of strings that have resulted
from the permutation of the $X^\mu_L$-mover degrees of freedom it
could (in some cases, with sufficient delay) be interpreted as the
scattering of the first pair of strings.
But if it occurred in the same moment of time in some frame it does not seem
easy to interpret them this way.

Another idea that could help  on the problem is to make use of the
gauge-like transformation (\ref{transf})

\begin{eqnarray}  
X^\mu_R\rightarrow X^\mu_R+k^\mu\\\nonumber
X^\mu_L\rightarrow X^\mu_L-k^\mu
\end{eqnarray}  

which can be used on the right and left constituents contributing to a 
single string.
For another string you may choose another 26-vector for $k^\mu$.
Considering two sets of this type of transformation for two (say
incoming) strings, the cross combined strings will not have their
positions invariant but will rather be displaced by
$\pm(k^\mu_{(1)}-k^\mu_{(2)})$ where $k^\mu_{(1)}$ and $k^\mu_{(2)}$
are the shiftings for the first and the second string respectively.
If one somehow thinks of these $k^\mu_{(1)}$ and  $k^\mu_{(2)}$ as
random and spreading over all 26-space we would almost certainly get
the cross combined strings out of sight in practice.

In this philosophy one expects scattering amplitudes (=$S$-matrix
elements) to be computed basically by putting up the state in our
Fock-space(s) [both $H_R$ and $H_L$ in the only closed string case and
the identified space $H_R=H_L$ in the open string case] corresponding
to the incoming strings and then simply take the Hilbert inner product 
overlap with the corresponding outgoing system

\begin{eqnarray}
\left<\mathrm{outgoing}|S|\mathrm{ingoing}\right>=
\left<
\begin{array}{c|c}
\mathrm{outgoing}&\mathrm{ingoing}\\ 
\mathrm{in ~ our  ~ string ~ field ~ theory}&\mathrm{in ~
our ~ string ~ field ~ theory}%\\ 
%\mathrm{in ~ our ~ string ~ field ~ theory}&\mathrm{ingoing ~ in ~
%our ~ string ~ field ~ theory} 
\end{array}
\right>
\end{eqnarray}

The scattering so to speak is immaterial and nothing really happens in
our string field theory. 
It is totally ``free'', scattering is all phantasy!

\section{The successor operator and the commutation rules of $X^\mu_R$ 
with itself}

There can be considered to be two motivations for imposing the
condition mentioned under step 2) in the foregoing section:

a) we like the constituents to form chains/curves in $X^\mu_R$-space
(or $X^\mu_L$-space).

b) we must implement a restriction corresponding to the feature 

\begin{equation}
\Pi^\mu_R=\frac{d}{d\tau_R}X^\mu_R\cdot\frac{1}{\pi\alpha'}
\label{RLC}
\end{equation}

which occurs for a physically reasonable assignment of the momentum
density $\Pi^\mu$ to be a sum $\Pi^\mu_L+\Pi^\mu_R$ of terms
associated with the left- and right-mover degrees of freedom.
This relation (\ref{RLC}) really tells us that the $X^\mu_R$-degrees of 
freedom are only one half degrees of freedom for each dimension, as
can also be seen from the commutation relation

\begin{equation}
\left[
X^\mu_R(\tau_R), X^\nu_R(\tau'_R)
\right]
=-ig^{\mu\nu}\theta(\tau_R-\tau'_R)
\end{equation}

which shows that the $X^\mu_R$ for different $\tau_R$'s do not
commute.

Translated into the language of constituents we must think of 
a discrete but very dense chain of them, each constituent covering (so 
to speak) a very small interval in the $\tau_R$-variable say, of length 
$\triangle\tau_R$.
Then we should identify the momentum $p^\mu$ of the constituent at 
$\tau_R$ with

\begin{equation}
p^\mu=\triangle\tau_R\Pi^\mu_R(\tau_R)\cdot\frac12
\end{equation}

(the $\frac12$ comes because of our normalization of $\Pi^\mu_R$ so
that $[\Pi^\mu_R, X^\nu_R]=-i\frac12g^{\mu\nu}\delta(\sigma'-\sigma)$)

On the other hand

\begin{equation}
-X'^\mu_R(\tau-\sigma)=\dot X^\mu_R(\tau-\sigma)=\frac{d}{d\tau_R}X^\mu_R(\tau_R)
\end{equation}

also means the differentiation as you go along in $\tau_R$ and
therefore $\frac{1}{\triangle\tau_R}\cdot$ (the step between
constituents) $\approx\frac{X^\mu_R(i+1)-X^\mu_R(i)}{\triangle\tau_R}$ 
where $i$ is the number of a constituent along the chain.
We must thus have 

\begin{equation}
\frac{X^\mu_R\left(\stackrel{i+1}{\tau_R+\triangle\tau_R}\right)-X^\mu_R(\stackrel{i}{\tau_R})}{\triangle\tau_R}=\Pi^\mu_R\cdot2\pi\alpha'
=\frac{2p^\mu2\pi\alpha'}{\triangle\tau_R}
\end{equation}

and thus we have 

\begin{equation}
X^\mu_R(i+1)-X^\mu_R(i)=4p^\mu\pi\alpha'
\end{equation}

Here the ``successor'' of the $i$th constituent is the $(i+1)$th and
we may seek to construct an operator $O$ that can bring the state
$\psi_i$ of the $i$th constituent into that for the successor, the
$i+1$th, 

\begin{equation}
\exp\left(
i\frac{\pi\alpha'\cdot 4}{2}p^2
\right)
=\exp\left(
i2\pi\alpha'p^2
\right)
\label{fsucc}
\end{equation}

To get the factors 2 right here we should be careful with the
commutation rules in our notation 

\begin{eqnarray}
\Pi^\mu=\Pi^\mu_R+\Pi^\mu_L\\\nonumber
X^\mu=X^\mu_R+X^\mu_L
\end{eqnarray}

and 

\begin{equation}
\left[
\Pi^\mu(\sigma,\tau), X^\nu(\sigma', \tau)
\right]
=-ig^{\mu\nu}\delta(\sigma-\sigma')
\end{equation}

and thus using also $[X^\mu_R, X^\nu_L]=0$, $[\Pi^\mu_R, \Pi^\nu_L]=0$ having

\begin{equation}
\left[
\Pi^\mu_R, X^\nu_R
\right]=-i\frac12g^{\mu\nu}\delta(\sigma-\sigma')
\end{equation}

The operator (\ref{fsucc}) will do the job of creating a state
displaced by $4\pi\alpha'p^\mu$ in $X^\mu_R$-space, but any operator of
exponential form with a function of the $X^\mu_R$-operator, say
$f(X^\mu_R)$ added will do this job too.
The appearance of such a freedom in the choice of the operator, which
could thus be

\begin{equation}
\exp\left(
i2\pi\alpha'p^2+f(X)
\right)
\end{equation}

for each coordinate in $X^\mu_R$-space or generally formulated

\begin{equation}
\exp\left(
i2\pi\alpha'p_\mu p^\mu+f(X^1_R, X^2_R, \cdot\cdot\cdot X^{25}_R, X^O_R)
\right)
\end{equation}

is welcome and not unexpected.
Indeed there should be what we called ``half a degree of freedom'' per 
point in choosing how the chain of constituents should be embedded
into $X^\mu_R$-space.  
In the classical
approximation there should be one real parameter to determine the  
state of one constituent, once the state of the foregoing is known. 
We thus expect the operator $O=\exp\left(
i2\pi\alpha'p^2+f(X_R)\right)$ to have one real free parameter with
which implement this ``half'' a degree of freedom in the classical
approximation.
In classical approximation we consider the constituents in wave packet 
states with so small extension in both $X^\mu_R$-space and its conjugate
that we can consider interesting functions as slowly varying over such 
small distances.
So in this classical approximation we would like to Taylor expand the
function $f(X^\mu_R)$ and approximate it by its term linear in $X^\mu_R$
and that would just allow one real parameter to use to parameterize
the state of the chain of constituents.
It could be tempting and fun -- but it ought not to be of much
importance in classical limit -- to play with the second order term in the 
Taylor expansion of $f(X^\mu_R)$ and put it in ``for beauty'' with a small 
once for all settled coefficient.  
Then we can combine it with the
$2\pi\alpha'p^2$ term and make out of the whole exponent a harmonic
oscillator Hamiltonian multiplied by the imaginary unit $i$.
We then propose for elegance -- hoping that the details do not 
matter much -- the operator $O$ of the form 

\begin{equation}
O=\exp(iH)
\end{equation}

where $H$ is the harmonic oscillator Hamiltonian with the kinetic term 
enforced to be of the form $2\pi\alpha'p^2$.  The mass
$M_{\mathrm{osc}}$ of the
oscillating particle is given by 

\begin{eqnarray}
\frac{1}{2M_{\mathrm{osc}}}=2\pi\alpha'\\\nonumber
i.e. ~ M_{\mathrm{osc}}=1/(4\pi\alpha')
\end{eqnarray}

Therefore the Hamiltonian is given by

\begin{equation}
H=2\pi\alpha'p^2+\frac12K(X^\mu_R-X^\mu_{Ro})^2
\end{equation}

The frequency of this formal oscillator is 

\begin{equation}
\omega=\sqrt{\frac{K}{M_{\mathrm{osc}}}}=\sqrt{4\pi\alpha'K}
\end{equation}

and for the use of the ``Hamiltonian'' $H$ in the exponent $\exp(iH)$
the dimensions shall be so that $\omega \; \hbar$ is dimensionless, or with
$\hbar=1$ $\omega$ should be dimensionless.
That is to say the dimensions should be 

\begin{eqnarray}
\left[M_{\mathrm{osc}}\right]=
\left[\frac{1}{4\pi\alpha'}\right]=
\left[GeV^2\right]\\\nonumber
\left[K\right]=
\left[\frac{1}{\hbar^2}\cdot\frac{1}{4\pi\alpha'}\right]=
\left[GeV^2\right]
\end{eqnarray}

We could for instance choose $K$ so that $\omega \; \hbar$ the spacing between
levels would be $2\pi$ divided by some large natural number, $q$ say,
so that the $\hat O=e^{-i\frac12\omega \; \hbar}O$ deviating only from $O$ by a 
constant phase factor would have the property 

\begin{equation}
\hat O^q=1
\end{equation}

Only in the limit $q\rightarrow\infty$ we would really get the string
theory, but it would be interesting to see if this special idea could
be relevant to connect to the string theories with $q$-adic numbers if 
$q$ were a prime.
We would obtain $\omega \; \hbar=\hbar\sqrt{K4\pi\alpha'}=\frac1q$ for
$K=\frac{1}{q^24\pi\alpha'}$. 
The parameter $X^\mu_{Ro}$ that denotes the bottom of the potential for the
(analogue) oscillator is the one that gives the ``half'' degree of
freedom.  

Now the application of the operator $O$ or $\hat O$ should be that
the allowed subspace of the Fock-Hilbert space for the $X^\mu_R$-space
constituents $H_{R ~ \mathrm{allowed}}\subseteq H_R$ is built up by
inserting products of a large (in principle in the  infinitely
many limit) number of creation operators, each creating the particles in a
state connected by $O$ to the foregoing state.
The states in $H_{R ~ \mathrm{allowed}}$ are
constructed from a ``vacuum'' state in $H_R$ by action with products
of creation operators of the form 

\begin{equation}
\Pi_ka^+\left(\circ\hspace{-0.27cm}\Pi^k_{l=0}\hat O_l(X^\mu_{Rol})\Psi\right)
\label{chainop}
\end{equation}

Here the state of the $k$th creation operator
$a^+\left(\circ\hspace{-0.26cm}\Pi^k_{l=0}\hat
O_l(X^\mu_{Rol})\Psi\right)$, which we could call

\begin{equation}
\Psi_R=\circ\hspace{-0.25cm}\Pi^k_{l=0}\hat
O_l(X^\mu_{Rol})\Psi
\end{equation}

is obtained from a starting state $\Psi$ by a series of successive
applications of the ``going to the successor'' operator $\hat
O_l(X^\mu_{Rol})$. Here a bottom of the oscillator point 
$X^\mu_{Rol}$ varies -- in a smooth way -- as one goes along 
with the chain-link enumerating integer $l$.
Also $k$ is used to enumerate constituents along the chain.
The product sign $\circ\hspace{-0.26cm}\Pi$ means product
with respect to the function composition $\circ$.  But the
functions $\hat O_l(X^\mu_{Rol})$ composed are really just linear 
operators acting on the single particle Hilbert space so that we would 
really denote usually the product without using the function
composition sign $\circ$.
To start the chain at some point is not allowed.
It is meant that either $k$ must run infinitely in both positive and
negative directions -- so that it should be made sense also of $k$
being negative -- or it should make up a closed loop chain.

It is of importance for the appearance of scattering at all, even if
it gets somehow phantasy only in our model, that the same state in the 
allowed Fock space $H_{R ~ \mathrm{allowed}}$ may be created by
several different combinations of chain operators (\ref{chainop}).
At least it is of importance that states made by different chain
combinations may have a nonzero overlap.
Otherwise there could not even in some point of view -- of
``phantasy'' -- be any scattering.
But that is also possible since two different chains could
even have a constituent each so that these two constituents would be
in exactly the same single particle state.
Then one could make an ``overlapping'' pair of chains to a given
two-chain state where the two chains have such a common constituent
state in this way: take a new pair obtained by tail exchange from the
first pair.  
Thus we can construct another two-string state in which
compared to the first pair the strings are constructed from one
``half'' from each string in the first pair.
The switch over between the two interpretations could be done at the
common state constituents.
Note how the switching of tails lacks of significance which was
announced to characterize our type of string field theory.  
It is in the just 
given example used to argue that we could make examples of the same
Fock state of e.g. two chains created on it, being
constructable in more than one way.

You may also bear in mind that this property of our allowed Fock space, 
which has same state constructable in several ways, i.e. from several
chain (=``half strings'') combinations, is the one that allows
scattering to seemingly take place without anything having to happen in 
the Fock space language in our model.
It is the property that scattering is allowed to take
place as pure phantasy.

\section{Idea of Deriving the Veneziano Model}

Since it has now been suggested that fake scattering should be
possible, it is very important to see if such ``fake''
scattering indeed will go according to the Veneziano model as its
scattering amplitude.
Otherwise our string field theory would not correspond to the string
theory.

We shall limit ourselves to a sketch of an argument only by suggesting
that at least it can look likely that the Veneziano model will result.

What we have to do in order to look for a Veneziano model scattering
amplitude is to think of a number of incoming strings -- in some mass
eigenstates say $|1, p_1; 2, p_2;
\cdot\cdot\cdot; q, p_q>$ and then look 
for overlap of this state with the outgoing state $|q+1,
-p_{q+1}; \cdot\cdot\cdot; N, -p_{_N}>$.
This overlap should then hopefully turn out to be the Veneziano model
scattering amplitude.

The main idea in the derivation of the Veneziano model as we hope to
perform it from our model is that the ``summation'' over the different 
ways of combining ingoing and outgoing particle constituents becomes
the integration in the Veneziano model expressions.
We can say that this
integration/summation over various assignments of ingoing to outgoing
constituents roughly speaking should become the Koba-Nielsen variables. 

Although we have not yet completed the calculation\cite{nielsen}, it would be
strange if we did not get Veneziano model when calculating the
overlaps proposed.

\section{Conclusion and resume}

We have put forward ideas for second quantizing string
theory into a string field theory in a way that is suggested to be
probably different from string field theory schemes in the cases of
Kaku-Kikkawa, Kyoto group, the cubic theory of Witten and Zwiebach.
Rather our formulation or model is in class of strings in QCD or
matrix model strings.
Indeed our model is in a class we could call ``constituent string field
theory.''
One of the consequences in this class is that two strings 
following each other for a piece in the middle (or just meeting
somewhere) cannot be distinguished from the string pair obtained by
tail-exchange.  
Furthermore our string field theory is characterized by the fact that
it has right mover
and left mover degrees of freedom expressed by constituents
separately.
In the
case of theories with open strings the Hilbert space of all the
possible numbers of strings is suggested to be described by a
certain subspace -- the space of allowed states $H_{\mathrm{allowed}}$
-- of the
Fock space for massless (scalar in bosonic string case) particles.
This ``allowed'' subspace consists of state-vectors that can be
constructed by operators creating chains of constituents.  One
constituent has its state obtained from the foregoing one (along the chain) 
by action of an operator with a parameter.  Thus the freedom of
inserting such chains corresponds so to speak to ``half a degree of
freedom per constituent''.
In this way the model gets the right number of degrees of freedom since 
there is locally both $X^\mu_R$ and $X^\mu_L$ which makes up
$X^\mu=X^\mu_L+X^\mu_R$ although in the open string globally $X^\mu_R$ 
and $X^\mu_L$ are constructed from the same constituents.

It is to be stressed that in our string field theory the scattering,
meaning that strings exchange parts with each other as time passes is not
represented as anything happening in the fundamental language of our
model!
That is to say that in our model -- fundamentally speaking -- no
scattering happens; it is rather just a fake.
The point is that we have declared the features of the multistring
states which change under a scattering process for non-existing as
fundamental degrees of freedom.
We do not distinguish the scattered and yet unscattered
systems of strings.
This may seem at first to look like throwing too much information out
of our scheme but we suggest that it should nevertheless be possible
to obtain the Veneziano model as the transition amplitude between the
incoming and outgoing multistring states in the Fock space in our
model.
If this is confirmed it will support the thesis that our model is
indeed a description of string theory.
Remarkably enough our model is really a free particle Fock space
theory.

So from our model point of view the hypothesis is that Nature should be
described by string theory would raise the question: Could Nature
indeed be successfully described by a free theory?
Presumably it can, but one might be worried how a Thuring machine
could be embeded in a totally free theory.
How could all the complicated computations be done by a machine
finally described by a ``free'' model?

What would be helpful in making the working of our scheme more
trustworthy would be if we find how some of the features of string
theory like branes, favoured dimensions and gauge groups, are to be
seen in our scheme.
Since our model is in many ways much simpler -- really a free
massless particle theory, with some constraints on the allowed states 
though -- one could hope that such feature, D-brane, dimension,
etc., might be seen rather differently and perhaps more
easily.

\end{document}

\end{document}